\documentclass{pasa}%
\usepackage{graphicx,caption,subcaption,color}
\newcommand{\feh}{\mbox{$\rm [Fe/H]$}}

\newcommand{\alphafe}{\mbox{$\rm [\alpha/Fe]$}}

\newcommand{\rgc}{\mbox{$\rm R_{GC}$}}

\newcommand{\aap}{A\&A}
\newcommand{\apj}{ApJ}
\newcommand{\apjl}{ApJL}                                     

\newcommand{\mnras}{MNRAS}
\newcommand{\aj}{AJ}

\usepackage[authoryear]{natbib}
\usepackage{color}
\bibpunct{(}{)}{;}{a}{}{,}
\title[The Metallicity Distribution of the Milky Way Bulge]{The Metallicity Distribution of the Milky Way Bulge}
\author[M.~Ness \& K.~Freeman]{M.~Ness$^1$, K.~ Freeman$^2$\\
\affil{$^1$Max-Planck-Institut f\"ur Astronomie, K\"onigstuhl 17, D-69117 Heidelberg, Germany}%
\affil{$^2$Research School of Astronomy and Astrophysics, Australian National University, Cotter Rd., Weston, ACT 2611, Australia}}%
\jid{PASA}
\doi{10.1017/pas.\the\year.xxx}
\jyear{\the\year}

\begin{document}%
\begin{abstract}%

The Galactic bulge of the Milky Way is made up of stars with a broad range of metallicity, --3.0 $<$ [Fe/H] $<$ 1 dex. The mean of the Metallicity Distribution Function (MDF) decreases as a function of height $z$ from the plane and, more weakly, with galactic radius \rgc. The most metal rich stars in the inner Galaxy are concentrated to the plane and the more metal poor stars are found predominantly further from the plane, with an overall vertical gradient in the mean of the MDF of about ---0.45 dex/kpc. This vertical gradient is believed to reflect the changing contribution with height of different populations in the inner-most region of the Galaxy. The more metal rich stars of the bulge are part of the boxy/peanut structure and comprise stars in orbits which trace out the underlying X-shape. There is still a lack of consensus on the origin of the metal poor stars ([Fe/H] $< -0.5$) in the region of the bulge.  Some studies attribute the more metal poor stars of the bulge to the thick disk and stellar halo that are present in the inner region, and other studies propose that the metal poor stars are a distinct `old spheroid' bulge population. Understanding the origin of the populations that make up the MDF of the bulge, and identifying if there is a unique bulge population which has formed separately from the disk and halo, has important consequences for identifying the relevant processes in the the formation and evolution of the Milky Way. 

\end{abstract}

\maketitle%

\section{The global MDF of the bulge} 

\begin{figure}[h!]
\begin{center}
\includegraphics[totalheight=0.40\textheight]{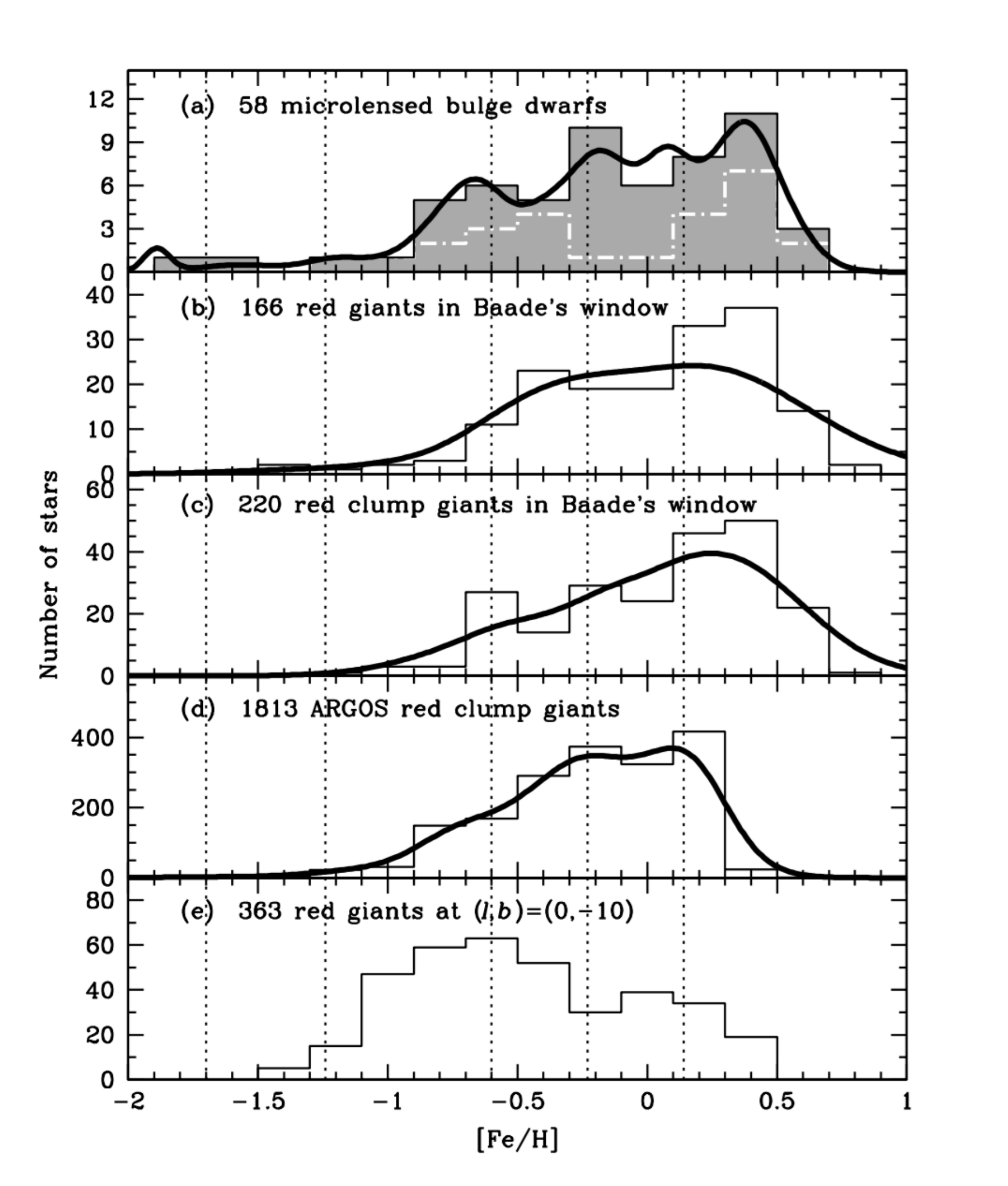}
\caption{Bulge MDFs from several sources assembled by Bensby et al., 2013 showing:  a) The MDF for the microlensed dwarf sample (the white dashed line shows the 26 microlensed dwarf stars from \citet{Bensby2011}; b) 166 red giant stars in Baade$^\prime$s window from \citet{Hill2011} c) 220 red clump stars in Baade$^\prime$s window from \citet{Hill2011} d) 1813 red giant stars from the ARGOS survey fields at (l, b) = (0$^\circ$, --5$^\circ$), (5$^\circ$, --5$^\circ$), (--5$^\circ$,--5$^\circ$) from \citet{Ness2013a}; e) 363 red giants at (l, b) = (0, --10$^\circ$) from \citet{Uttenthaler2012}; The curves are generalised histograms and the dotted lines mark the peaks of the MDF components proposed by \citet{Ness2013a}.}
\label{fig:thomas}
\end{center}
  \end{figure}

The stars of the bulge of the Milky Way have a broad metallicity distribution,  with [Fe/H] between $-3.0$ and  $+1$ dex. Spectroscopic surveys of the bulge, both high and medium resolution,  have reported  a metallicity gradient in latitude. This was first measured to be about $-0.35$ dex kpc$^{-1}$ along the minor axis of the
bulge by Minniti et al (1995), and then $-0.6$ dex kpc$^{-1}$ by \citet{Zoccali2008} from observations of about 900 K giants across $b = -4^\circ$ to $b = -12^\circ$ on the minor axis. \citet{Ness2013a} later found a similar minor axis gradient of $-0.45$ dex kpc$^{-1}$ from spectroscopy of about 2000 red clump giants with \rgc\ $< 3.5$ kpc,  across $b = -5^\circ$ to $b = -10^\circ$.  Numerous studies \citep[e.g.][]{Hill2011, Babusiaux2010, Gonzalez2011, Uttenthaler2012, RJ2014, Johnson2012, Ness2013a, Bensby2013} have reported similar MDFs for the bulge.  Figure \ref{fig:thomas}, from \citet{Bensby2013} compares the MDFs from a number of different studies in various bulge fields.   In addition to the vertical metallicity gradient, there is also a weaker gradient in longitude, at $b > 4^\circ$. This is seen in the photometric metallicity distribution by \citet{Gonzalez2013}. The vertical abundance gradient was initially interpreted as inconsistent with the formation of the bulge from the disk via dynamical instabilities, which results in a boxy/peanut morphology similar to that already observed in the Milky Way \citep{Dwek1995}. This large abundance gradient was instead interpreted as evidence for a bulge formed via mergers or dissipational collapse \citep{Zoccali2008}. 

Subsequently, however, bulge metallicity gradients in $(l,b)$ like those measured for the Milky Way have been shown to be a natural consequence of bulge formation via instability of the disk, due to the preferential redistribution of stars according to their initial phase space energies in the disk \cite[][]{Inma2013,diMatteo2014}. Furthermore, the overall large negative gradient in the mean of the MDF is now believed simply to reflect the changing contribution of populations which have different spatial distributions in the innermost regions. The more metal-rich population of stars is concentrated to the plane, while the more metal-poor populations are present in greater number at larger heights from the plane. Although the various observers find very similar MDFs (see Figure \ref{fig:thomas}), there is no consensus yet in the interpretation of this MDF (see Section \ref{sec:pops}). 

\begin{figure*}
    \begin{subfigure}[b]{0.5\textwidth}
    \centering
       \includegraphics[scale=0.2]{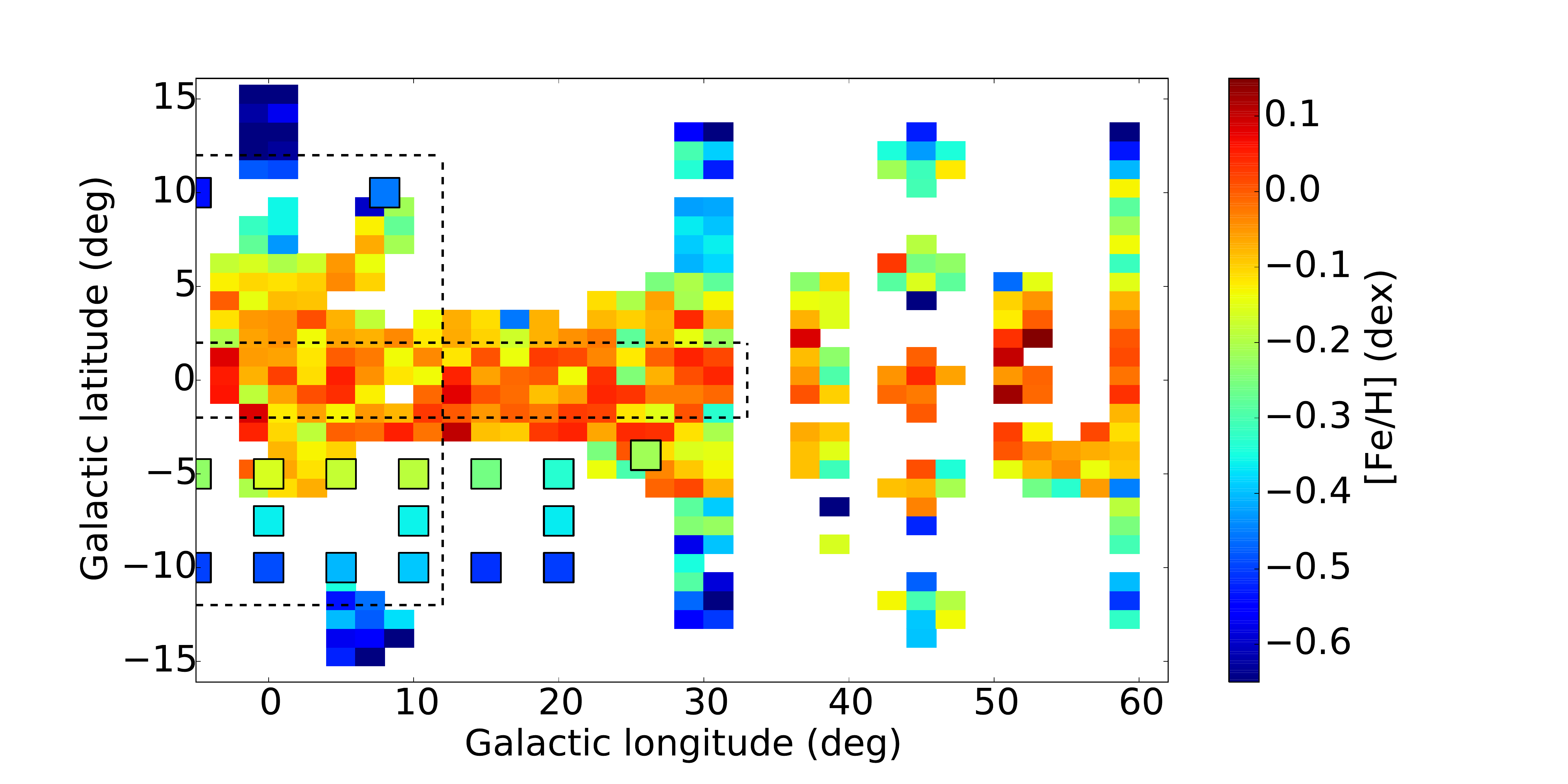}
\caption{\feh\ map}
  \end{subfigure}%
  \begin{subfigure}[b]{0.5\textwidth}
    \centering
        \includegraphics[scale=0.2]{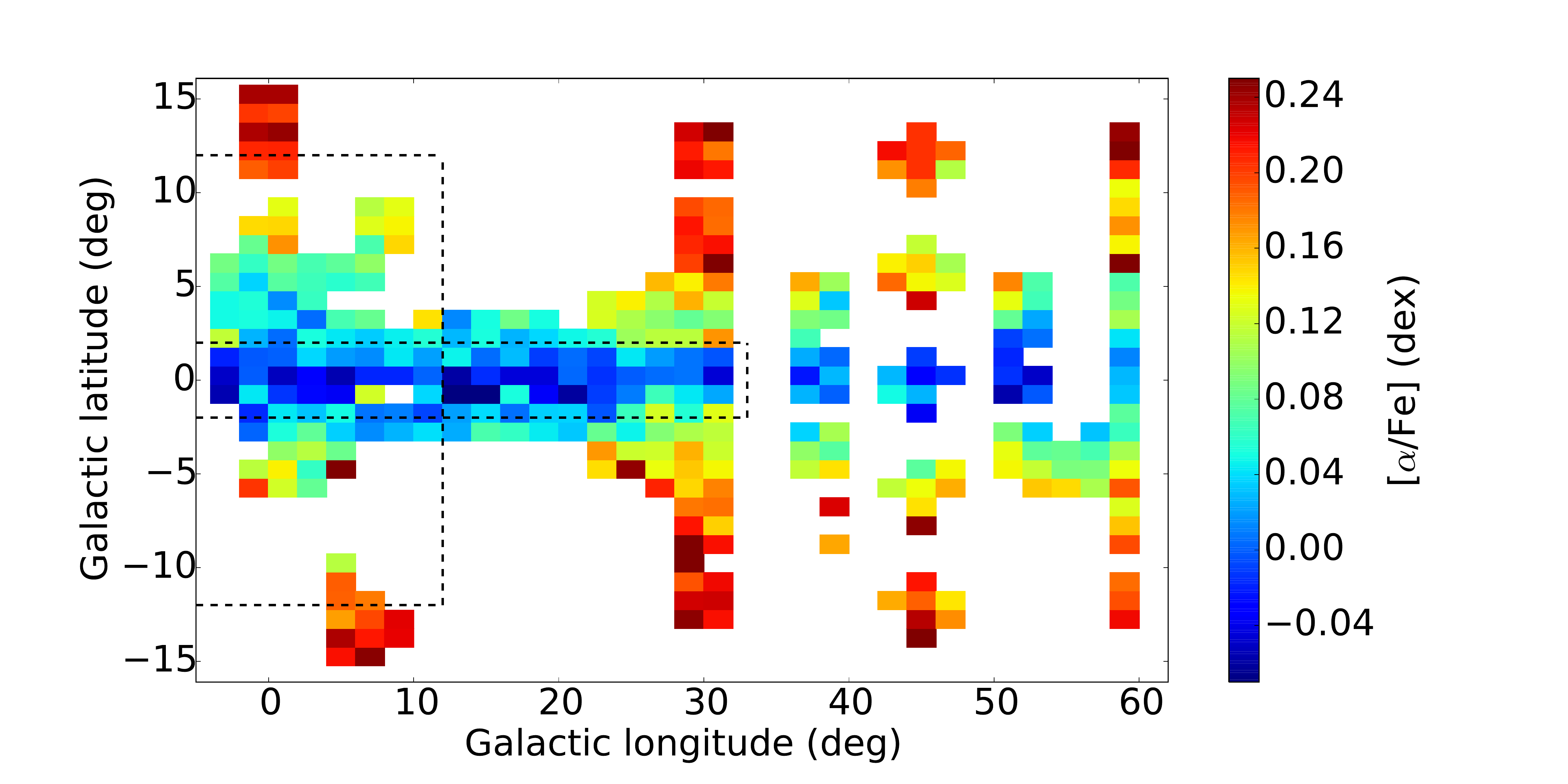} 
\caption{\alphafe\ map} 
\end{subfigure}
  \caption{(a) \feh\ and (b) \alphafe\ maps for the 8500 bulge and disk stars from APOGEE and 8000 ARGOS bulge stars (in the larger outlined boxes for the \feh\ map only) spanning heliocentric distances of 4--12  kpc. The dashed line indicates approximate outline  of the 180~pc thin bar identified by \citet{Wegg2015} and the larger box represents the approximate outline of the boxy bulge in the COBE image \citep{Dwek1995}.}
  \label{fig:apogee1}
\end{figure*}

 The vertical and (lesser) radial metallicity gradients reported for the MDF in the bulge by numerous studies have been found to flatten for latitudes $|b| < 5^\circ$. This was first reported from studies using small samples of stars near the minor axis \citep{Ramirez2000, Rich2007, Rich2012}. From the APOGEE survey, it has been revealed that the flattening of the gradient near the plane is seen right across the bulge in longitude,  out to $l < 10^\circ$ and not just on the minor axis, again for latitudes $< 5^\circ$.  This is shown in Figure \ref{fig:apogee1} which includes 8500 stars from the APOGEE survey and 8,000 stars from the ARGOS survey (those at positive longitudes where APOGEE has observed). These stars are at distances from $4$ to $12$ kpc from the Sun, thus showing stars in the bulge region (and stars in the disk away from the Sun to illustrate the smooth transition between the bulge and the disk). The mean \feh\ of these stars is shown in Figure \ref{fig:apogee1} (a) and Figure \ref{fig:apogee1} (b) shows the mean \alphafe.  The ARGOS fields, which each have about 600 stars, are shown in the larger boxes with black outlines for Figure (a) only. There are typically $\ge$ 30 stars in each of the bins which show the mean measurement for the APOGEE sample. 
 
From Figure \ref{fig:apogee1} (a) it is clear that in the inner region with (l,b) $< (10^\circ$, $5^\circ)$ the metallicity is nearly constant at constant latitude. The most metal-rich and correspondingly alpha-poor stars are concentrated to the layer $b < 2^\circ$, and these metal-rich  stars show a smooth transition out into the disk.  In the plane, the stars reach the highest metallicity along the thin long bar reported by Wegg et al. (2015), the profile of which is shown by the thin long box in Figure \ref{fig:apogee1}, extending out to $l = 30^\circ$. The outline of the COBE boxy/bulge profile is shown in this Figure as the larger box, going out to longitudes of $l \approx 10^\circ$. The metallicity measurements for $b > 4^\circ$ from APOGEE and ARGOS are in good agreement with the photometric metallicities of \citet{Gonzalez2013}

\section{Multiple Stellar Populations in the  Bulge} 
\label{sec:pops}

\begin{figure*}
\begin{center}
    \begin{subfigure}[b]{0.3\textwidth}
           \includegraphics[scale=0.2]{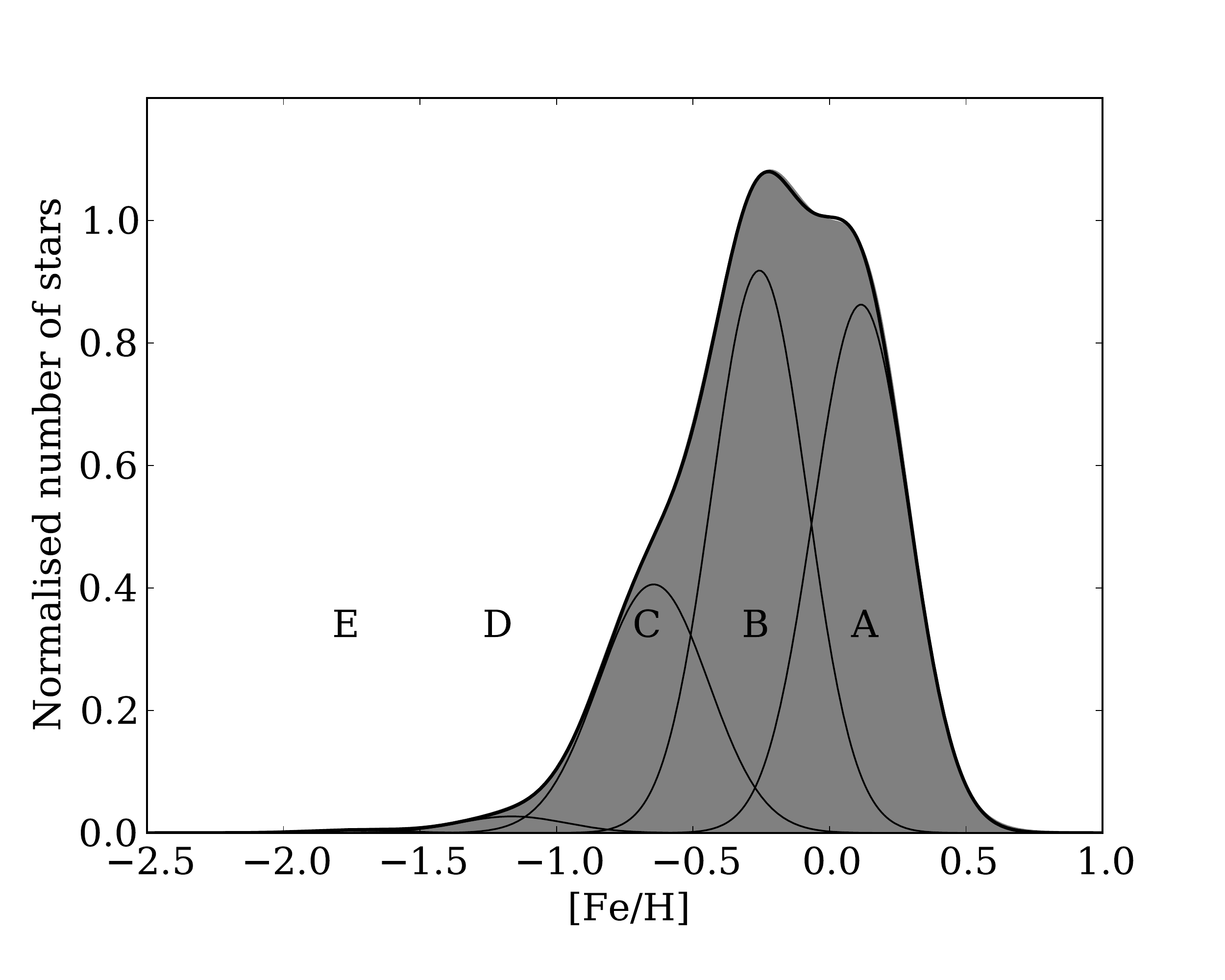}
        \caption{}
    \end{subfigure}
    \begin{subfigure}[b]{0.3\textwidth}
           \includegraphics[scale=0.2]{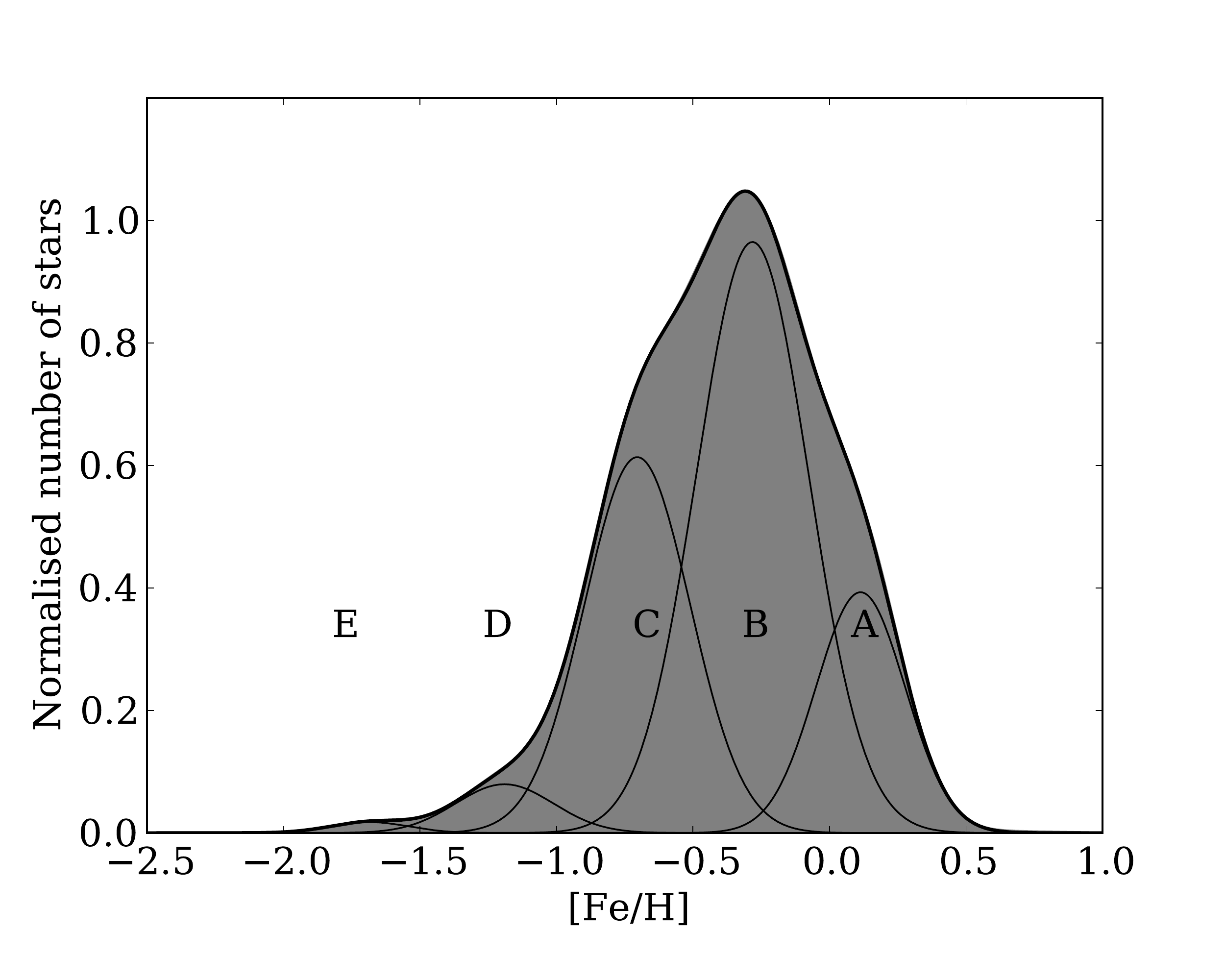}
        \caption{}
    \end{subfigure}
    \begin{subfigure}[b]{0.3\textwidth}
           \includegraphics[scale=0.2]{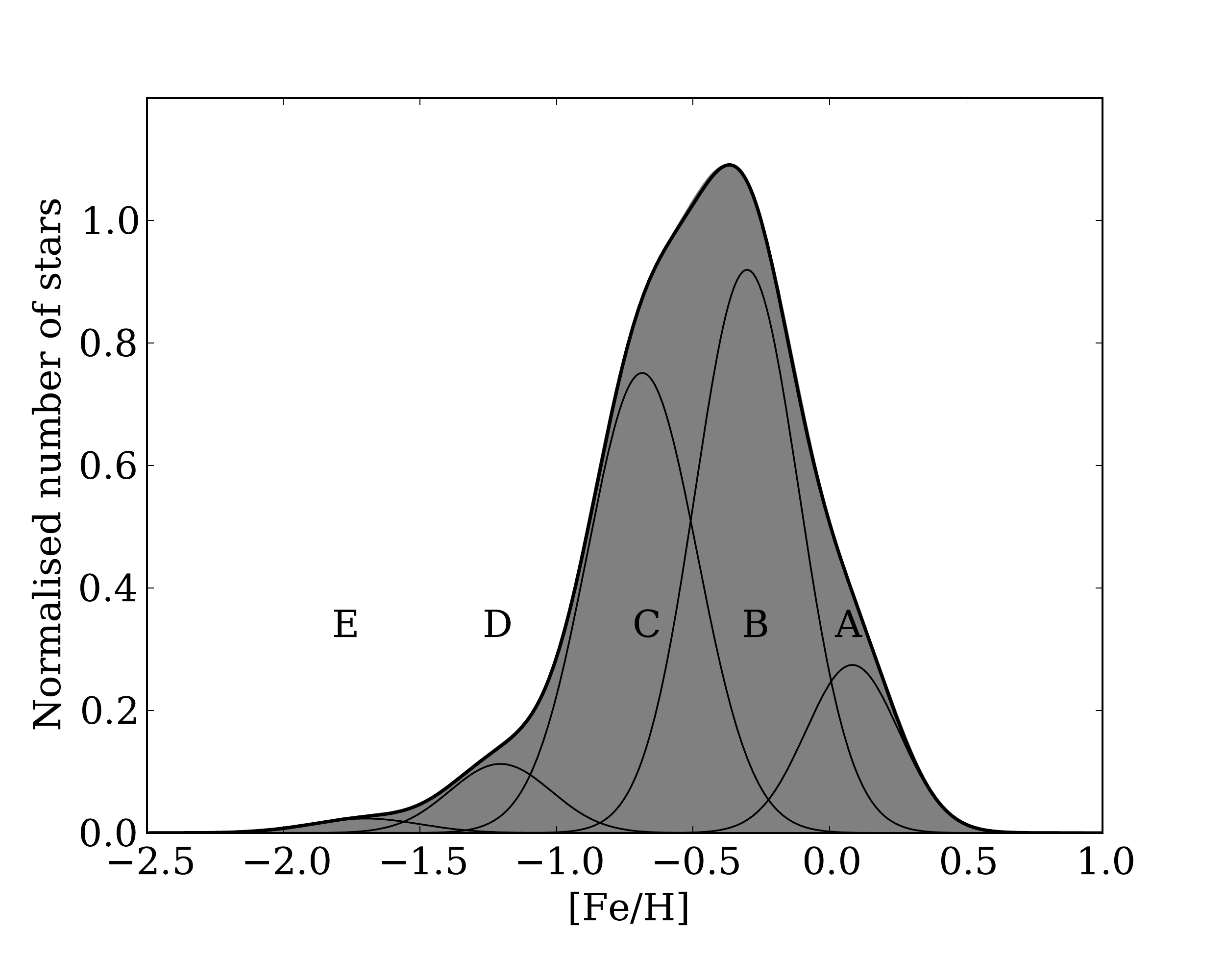}
        \caption{}
    \end{subfigure}
    \caption{The MDF for stars within \rgc\  $< 3.5$ kpc from the ARGOS survey: (a) for stars at $b = -5^\circ$, (b) stars at $b = -7.5^\circ$,  and (c) stars at $b = -10^\circ$, for all stars across longitudes $|l| < 15^\circ$ showing the changing contribution of metallicity fractions with latitude. The Gaussian components A-E are indicated.}
    \label{fig:Ness}
    \end{center}
\end{figure*}

Several studies \citep[e.g.][]{Babusiaux2010, Hill2011, Gonzalez2011, Freeman2012, RJ2014} have interpreted the MDF of the bulge across a broad region in $(l,b)$ as being made up of multiple populations. \citet{Ness2013a} interpret the MDF as having 5 populations which are shown in Figure \ref{fig:Ness}: 3 dominant populations (A--C) with [Fe/H] $> -1$ and two minor metal-poor populations (D \& E). From the ARGOS bulge survey described in \citet{Ness2013a}, the three populations A-C with \feh\ $> -1$ have peaks at metallicities of about +0.15, -0.25, -0.7 dex, respectively and provide about 95\% of stars in the bulge.  They associate these populations with the stars of the boxy/peanut bulge (A and B), the thick disk (C), the metal weak thick disk (D) and stellar halo (E). They find that these populations are present in different proportions throughout the  inner region of the Milky Way.

Using red clump stars, \citet{Ness2012} and \citet{Uttenthaler2012} showed that only the more metal rich stars in populations A and B (i.e. stars with \feh\ $> -0.5$) are part of the split density distribution of stars in the innermost region that reflects the X-shaped morphology of the bulge. This implies that stars with \feh\ $< -0.5$ that are present in the inner region are not part of the boxy/peanut bulge morphology and are not on X-profile-supporting  x1 orbit families.  The most metal-rich stars show the largest minima between peaks in the K-magnitude distribution of stars and are the most strongly involved in the X-shape \citep{Ness2012}. Population A was found by \citet{Ness2013a} to be concentrated toward the plane and to be the thinner part of the boxy/peanut bulge. Population B corresponds to the vertically thicker stars in the bulge, with a similar contribution fraction across $b = 5$ to $10^\circ$. Population C  is not significantly involved in the split density distribution or boxy/peanut structure but transitions smoothly from the bulge to disk with longitude; it is identified with the inner thick disk. The two populations of stars (A and B) associated with the boxy/peanut structure have similar peaked velocity
dispersion profiles with longitude. The velocity dispersion profiles for the more metal-poor stars have a different shape \citep[e.g.][]{Shen2010, Ness2013b, Portail2015} (see below). The  5\% of stars in the bulge with metallicities \feh\ $< -1.0$ are chemically similar to stars of the metal-weak thick disk and halo near the Solar neighbourhood \citep[e.g.][]{AlvesBrito2010,Bensby2013} and were associated by \citet{Ness2013a} with these populations.

\begin{figure*}
\begin{center}
    \includegraphics[totalheight=0.40\textheight]{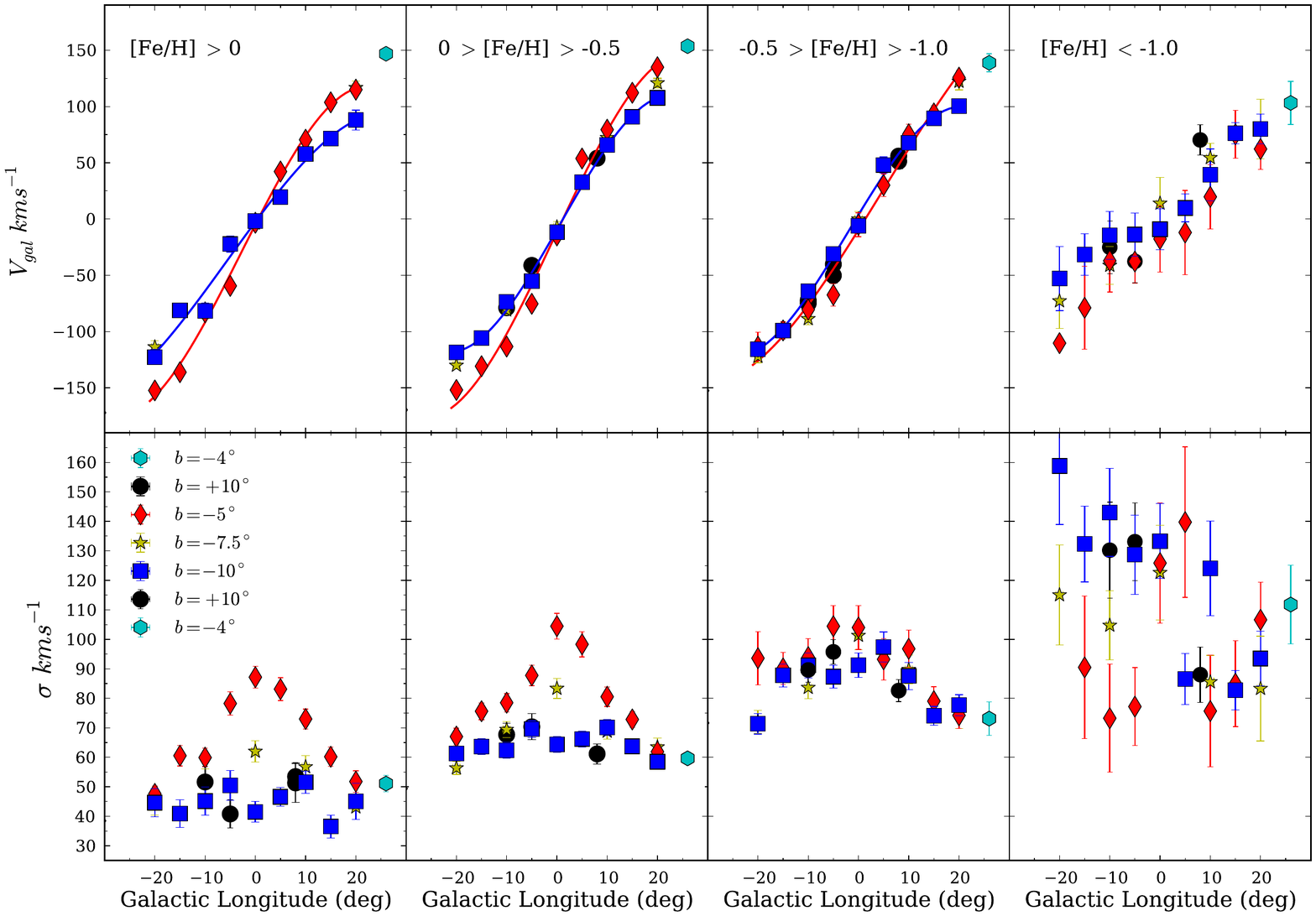}
\caption{The rotation curves (top) and dispersion profiles (bottom) for the 17,500 bulge stars from the ARGOS survey within distances 5 -- 11 kpc. The populations from most metal rich to most metal poor indicated in Figure \ref{fig:Ness} correspond to populations A,B,C and D/E. \citet{Ness2013b}.}
\label{fig:kinematics}
\end{center}
  \end{figure*}%
  The rotation and dispersion profiles reported by \citet{Ness2013b} and shown in Figure \ref{fig:kinematics} as a function of \feh\ support the differentiation of the populations. The kinematics of stars in populations A \& B are
related and are different from the other components. The stars in populations A and B, which are part of the split clump and the boxy/peanut bulge, show the same characteristic peaked pattern of velocity dispersion in the two left hand panels of Figure \ref{fig:kinematics}, with population A being a colder replica of population B. Population C,  
 associated with the thick disk in the inner Galaxy, is rotating as rapidly as the more metal-rich populations (all show the latitude-independent rotation profiles that are commonly seen in boxy bulges) but its velocity dispersion profile has a different shape. The most metal-poor stars with [Fe/H] $< -1$ have the slow rotation profile and high dispersion that might be expected of a population that had no ancestral association with a disk, such as a spheroidal population in the bulge or the stars of the inner halo or an underlying merger component. 

\begin{figure}
\begin{center}
    \includegraphics[totalheight=0.40\textheight]{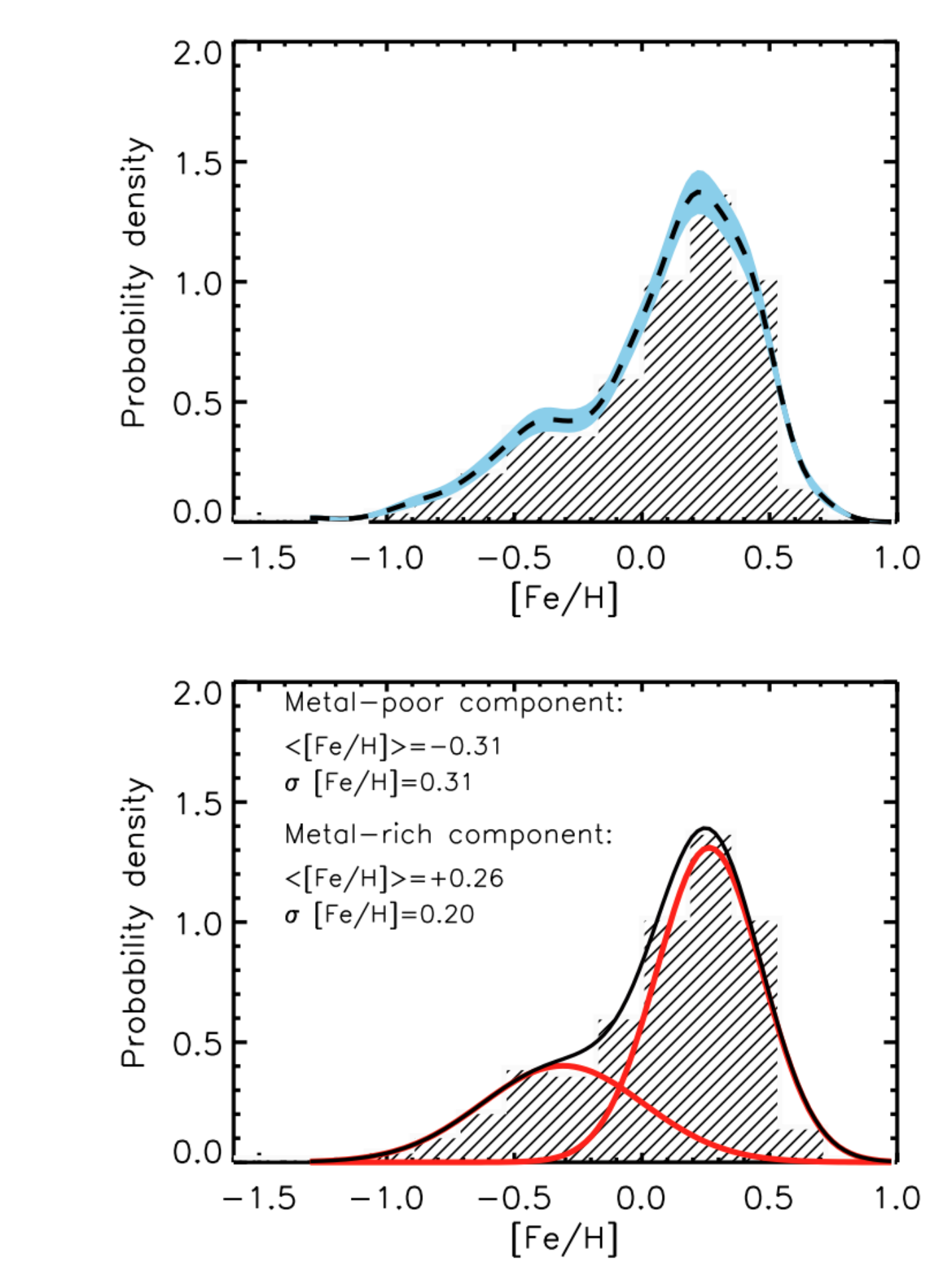}
\caption{From \citet{Gonzalez2015}: The MDF obtained from the combination
of four GIBS (high-resolution) fields and the red clump stars from
 \citet{Hill2011}. The probability density distribution is
shown in the upper panel as a dashed line, with the 
variance on the probability densities in
blue. The lower panel shows the best two Gaussian fit to the upper
panel distribution.}
\label{fig:gonzalez}
\end{center}
  \end{figure}

Other studies \citep{Babusiaux2010, Hill2011, Gonzalez2011, RJ2014} have proposed that the MDF of the bulge comprises two populations, a metal-rich population that is part of the boxy/peanut bulge and a metal poor population that is an old spheroid (i.e. with a possible formation history that is distinct from the disk and halo components of the Milky Way).  Although note that \citet{Gonzalez2015} concludes that whether their two MDF components have  different formation histories is yet to be determined.  The decomposition of stars into two roughly equal components from \citet{Gonzalez2015} is shown in Figure \ref{fig:gonzalez}, from the GIBS survey \citep{Zoccali2014}, where these two populations peak at [Fe/H] of about
+0.25 dex and -0.3 dex.  

The 2-component decomposition places fewer stars within the boxy/peanut or X-shaped structure, attributing approximately 50\% to an old spheroid \citep{Hill2011}. The 5-population model of \citet{Ness2013a} attributes 95\% of stars to be disk stars, with the population C (at \feh\ =  --0.7) being disk stars but simply not part of the X-shaped morphology (possibly because they were originally part of a hotter thick disk which was dynamically less responsive to the instability).  These two interpretations therefore have different implications for the stars in the inner region in terms of their origin from the disk and the contribution of any additional population that is distinct from any other Milky Way population and specifically from the bulge, such as an old spheroid formed via mergers at high redshift. Whereas a large fraction, up to 50\% of stars in the two-component decomposition, are associated with an old-spheroid component, only  $\sim$ 5\% of the population in the 5 population decomposition is associated with any old spheroid component, and this component may simply be metal-poor halo stars in the inner region and not a unique population of the bulge.

 \section{The alpha-enhancement of the bulge} 

The alpha-enhancement of stars in the bulge shows similar trends to that of the thin and thick disk in the solar neighbourhood, as reported by Alves Brito et al. (2013), Bensby et al. (2013) and Gonzalez et al.(2015). 

The APOGEE survey includes observations across the Milky Way's disk and bulge and allows the \alphafe\ trends to be homogeneously mapped from the outermost to the innermost region of the Galaxy \citep[see e.g.][]{Hayden2015, Nidever2014}. Comparing the \alphafe\ trends and the individual abundance patterns ([X/Fe]) for bulge stars with stars at other galactic radii is important for understanding the detailed formation of the bulge and its relationship with the disk. A large fraction of the bulge stars is likely to have come from the disk, and therefore the broad expectation would be that disk and bulge abundance trends will be similar. However, there are a number of perturbing processes that may show abundance signatures. These include the star formation rates in the inner and outer disk and the growth of the disk over time, gas inflow driving star formation, and radial migration. The alpha and abundance trends of the rare metal-poor stars in the inner Milky Way are particularly interesting due to the prediction that the oldest stars of the Galaxy will now be concentrated to the bulge \citep[e.g][]{brook2012, diemand2008, tumlinson2010}

\begin{figure*}
\begin{center}
   \includegraphics[scale=0.45]{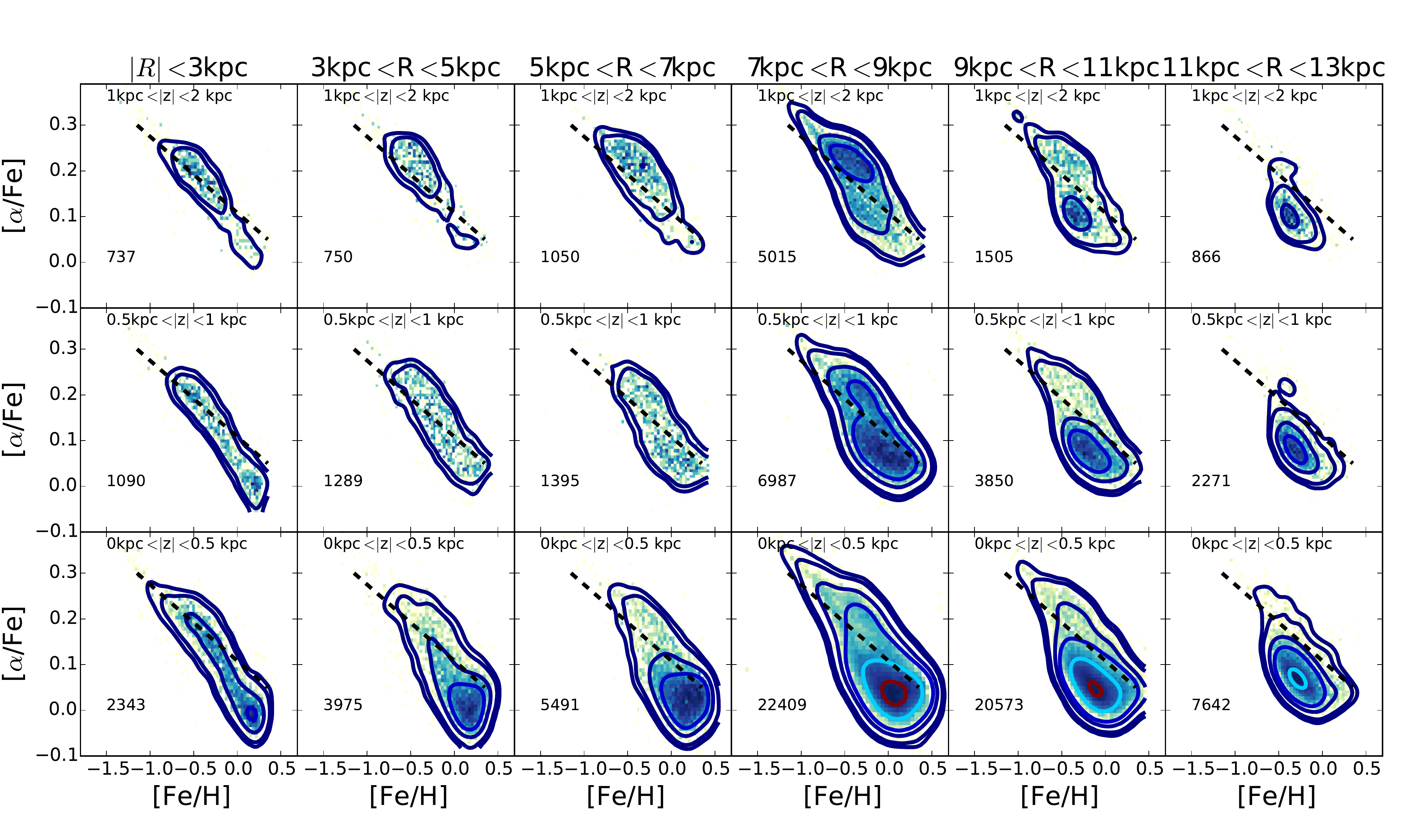}
  \caption{The \feh-\alphafe\ for the Milky Way from the Galactic center to \rgc\ = 13 kpc. Three intervals of height $z$  are shown, at increasing radii, for $0 < z < 0.5$ kpc, $0.5 < z < 1$ kpc and $1 < z < 2$ kpc, for the bottom, middle and top panels respectively. The number of stars is indicated in the left hand corner of each sub-panel. The dashed line shows the sequence in the inner bulge and is included to help follow the trends in the \alphafe\ sequence with increasing radius. Note that the alpha sequence narrows} at small galactic radii in the bulge, the low and high alpha sequences are both present near the sun, and only the low-alpha sequence is present at the largest radii in the disk. 
\label{fig:apogee_alpha}
\end{center}
\end{figure*}

Figure \ref{fig:apogee_alpha} shows the [Fe/H]--[$\alpha$/Fe] maps for the APOGEE stars for different Galactic zones in $R$ and $z$. The stellar parameters have been obtained with The Cannon \citep{Ness2015} and demonstrate that the inner Galaxy has the narrowest alpha sequence, which is offset to higher metallicity than in the outer galaxy.  The dotted line in Figure \ref{fig:apogee_alpha} marks the approximate trend of the high-alpha sequence that is seen in the innermost region. The density distribution shows that, in the zones further from the Galactic plane at a given radius, more stars are found in the low metallicity/high alpha sequence than in the high metallicity/low alpha sequence.  

Figure \ref{fig:apogee_alpha} shows that there is a smooth transition from the outer to inner region at a given height from the plane, with an increasing mean metallicity. The low-alpha sequence dominates in the outer region and a narrower sequence extending to lower \feh\ and higher \alphafe\ dominates in the inner region. At intermediate radii, near the Sun, both low and high alpha sequences are present.  The narrower and high-alpha sequence seen in the inner Galaxy implies that the star formation and chemical evolution rate was high at the early epoch in the disk at which the bulge stars were forming.  The star formation continued over a relatively brief period, and this single sequence is a snapshot in time of the rapid chemical evolution of the early disk from which the bulge was formed. \citet{Athanassoula2005} showed that bulges form early on after the disk, and over a short period of 1-2 Gyr. This rapid formation at early times is consistent with the single continuous alpha sequence seen in the inner 3kpc of the Milky Way that is observed in APOGEE data. The simulations of \citet{diMatteo2014} show that stars are mapped into the bulge from the disk according to their initial phase space location. Stars now in the bulge that reside nearest to the plane originate from the innermost region of the disk. Similarly, the stars in the bulge whose orbits take them to higher $z$ came from larger radii in the disk and preferentially at larger heights above the plane. Simulations that include star formation \citep[e.g.][]{obreja2013, calura2012} combined with new observational data e.g. from APOGEE which covers from the bulge and into the disk may offer insight into the observed chemodynamical distribution of stars across $(l,b)$.

\section{The metal poor \feh\ $<$ --1.0 population} 

About 5\% of the stars in the bulge have metallicities $<$ --1.0. The ARGOS survey showed that the fraction of these stars, which they associated with the Galactic halo population, increases with height above the plane. This population also rotates more slowly,  at about 50\% of the speed of more metal rich stars \citep{Ness2013b}. It also has a high and relatively latitude-independent velocity dispersion as seen in Fig. 4. \citet{Kunder2015} have also suggested that the halo population in the bulge is not insignificant, even within 1 kpc of the Galactic center. \citet{Bensby2013} measured the individual abundances of microlensed dwarf bulge stars and found that the stars with \feh $\approx$ --1.0 have very similar chemical abundance trends to the nearby thick disk stars. 

The bulge RR Lyrae population, whose MDF peaks around \feh $= -1.0$ \citep{Piet2014}, has been used to trace the distribution of the old metal-poor population of stars in the bulge. The RR Lyrae stars are ideal tracers of structure because their distances can be accurately estimated.  Although \citet{Dekany2013} reported the spatial distribution of the RR Lyrae to be spheroidal with only a slight elongation, the survey of \citet{Piet2014} showed that the metal-poor RR Lyrae very closely trace the barred structure of intermediate-age red clump giants.  \citet{Piet2014} show that the spatial distribution of the RR Lyrae population is barred but that these stars are not part of the X-shape structure. They also report two sequences of RR Lyrae with marginally different metallicities, proposing evidence for the role of mergers in the initial formation of the bulge.  Although they may follow the elongated distribution of the bar, the RR Lyrae stars may still be part of the halo but which has been pulled into a bar shape by the potential of the other stars \citep[e.g.][]{Saha2013}.

The high-resolution detailed chemical abundance analysis of a handful of the most metal-poor bulge stars with \feh\ $< -2.0$ by \citet{Howes2014} and \citet{GarciaPerez2013} reported similar individual element enhancement and alpha-enhancement trends to halo stars of the same metallicity (although \citet{Howes2014} noted an abundance scatter that may be larger). From the analysis of the dynamics and detailed chemical abundances of three bulge stars with \feh\ $< -2.7$, \citet{Casey2015} report that these stars follow the abundance trends identified previously for metal-poor halo stars, except for scandium. Simulations \citep[e.g.][]{brook2007, diemand2008, tumlinson2010} predict that the oldest stars in the Galaxy will be concentrated to the bulge. These very metal poor stars found by \citet{Casey2015} may well be members of the oldest populations of stars in the Milky Way, formed at redshift $z > 10$. The abundance patterns of this rare stellar population which is concentrated to the bulge could provide critical insight into the composition of the universe at the highest redshift and the conditions at the beginning of the Milky Way's formation.

\section{Conclusions}

Numerous studies, including small programs and large surveys observing the bulge across a broad spatial extent (e.g. ARGOS, GIBS, VVV) have characterised the overall metallicity distribution of the bulge.  Until recently, most spectroscopic studies have looked off the plane, at $|b| > 4^\circ$. Although a small number of fields nearer to the plane have been observed \citep[e.g.][]{Zoccali2014}, with the near-IR APOGEE survey \citep{Majewski2012}, the MDF can be examined for the bulge and for regions extending out into the disk, at lower latitudes and in the plane across the entire bulge and into the disk.

Detailed analysis from spectroscopic surveys has also shown the trends of alpha and individual element abundances
for stars of the bulge, at all metallicities, to be comparable with those of disk and halo stars of similar metallicity in the local neighbourhood. 

Overall the stars of the bulge show similarities to corresponding stars in the nearby disk and halo populations, but there are some critical differences.  The [$\alpha$/Fe] behaviour of bulge stars shown in Fig. 6 illustrates
the rapid history of chemical evolution that occurred in the inner Galaxy, probably before the bulge structure was in place.  The anomalous scandium abundance of the most metal-poor bulge stars (\feh $\sim -3.0$) may indicate that these very metal-poor stars in the bulge may indeed be among the first stars to have form in the very early Galaxy.  

Coupling the stellar metallicity distribution data with the kinematics and morphological structure \citep{Ness2012, Wegg2013, Wegg2015, Portail2015} is key to understanding the relationship between the bulge and the other populations of the Milky Way that were present at the time that the bulge was formed. Specifically this is important for understanding what fraction of the bulge was formed from pre-existing disk material, and what is
the nature of the bulge stars that are not associated with the boxy bulge.  Are they simply stars of the disk
and inner halo, or is there a unique bulge population that is not related to the disk? A lower limit of $\approx$ 25\% on the mass fraction in the X-structure has been recently determined by \citet{Portail2015}.   Constraining the total fractional mass of stars in the X-shape is of great interest, as a check on the consistency of the fractional
contributions of the different MDF populations with their likely origins.


\end{document}